# Amorphous Ferrimagnets: an Ideal Host for Ultra-Small Skyrmions at Room Temperature

S. Joseph Poon and Chung Ting Ma

Department of Physics, University of Virginia, Charlottesville, Virginia 22904 USA

**Abstract**

Recently, magnetic skyrmion has emerged as an active topic of fundamental study and applications in magnetic materials research. Magnetic skyrmions are vortex-like spin excitations with topological protection and therefore are more robust to pinning compared with magnetic domain walls. We employ atomistic simulations to create room-temperature ultra-small Néel skyrmions in amorphous ferrimagnet. The fast propagation and low-dissipation dynamics of ultra-small ferrimagnetic skyrmions make them attractive for utilization as an alternative to domain walls in spin-based memory and logic devices.

**Introduction**

In the 1960s, Professor Theodore (Ted) Geballe and his colleagues pioneered the study of the magnetic moment of 3D transition metal solutes in noble metals and discovered the formation of giant moment in these material systems [1]. The seminal study by Professor Geballe laid the foundation for the fundamental research on magnetic materials and, equally important, also impacted the study of strong electronic correlation and orbital effects. To celebrate his centennial birthday, it is fitting to discuss a new type of magnetic phenomenon called magnetic skyrmion (Figure 1), a topological excitation in a magnetic system that could potentially lead to better spin-based memory and logic devices. Originally, skyrmion was named for a certain type of topological soliton invented by T. Skyrme, also in the 1960s, in the study of particle creation in quantum fields.

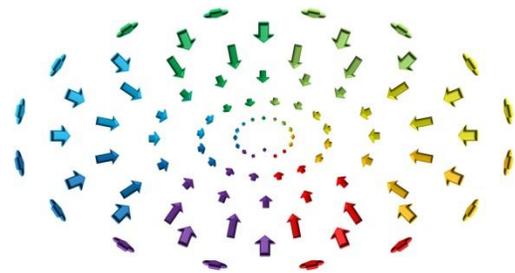

**Fig. 1** Top view of the spin texture in a Néel skyrmion from an atomistic simulation. Different from the domain wall spins, the spins in a skyrmion rotate by 360 degrees from one side to the other.

Beginning with some background, the discovery of giant magnetoresistance (GMR) led to the dawn of spintronics. Spintronics utilize the magnetic moment instead of electric charge to store data in magnetic memory, which potentially can have a significant impact on the future of electronics [2]. Non-volatile magnetic memory is one of the essential approaches in overcoming the von Neumann bottleneck in computer architecture. Current magnetic memory relies on the bubble and domain wall (DW) to encode information, using DWs in the race-track architecture to encode bits and to drive them with spin current. However, there are limitations that impede the advance of DW based magnetic memory. Skyrmions can be an alternative to DW race-track memory [3]. Being topologically protected and much smaller than DWs [4-5], e.g. ~10 nm or smaller in diameter, skyrmions can be easier to unpin from lattice defects. Along with the low driving threshold, magnetic skyrmions can form the basis for scalable and high-speed-low-power spin-based logic devices. However, such devices can only be possible if the small skyrmions of 10 nm or below in diameter exist sufficiently long at room temperature. In this article, we will employ atomistic simulations to study a class of materials, namely amorphous ferrimagnetic alloy films that are promising in hosting ultra-small skyrmions at room temperature. Our simulation study also showed stable ultra-small skyrmions at room temperature in ferromagnetic heterostructures, however, the conditions required are much more stringent.

**Method**

Skyrmions are stabilized via the Dzyaloshinskii Moriya interaction (DMI) [6-7], which originated from the interplay of spin-orbit effect and inherent chiral asymmetries or interfacial symmetry breaking. Figure 2 illustrates the chiral nature of the interfacial DMI. Intrinsic DMI arises in non-centrosymmetric crystals such as B20 alloy where Bloch skyrmions have been found at low temperature [8-9] Interfacial DMI originates from inversion symmetry breaking by a heavy metal interfacial layer such as Pt and Ir with strong spin-orbit coupling in multilayer stacks that contain ferromagnetic Fe and Co. The latter was found to host >40 nm Néel skyrmions at room temperature [10-12].

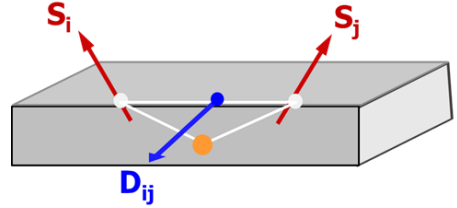

**Fig. 2** Schematic illustration of interfacial Dzyaloshinskii Moriya interaction. The DMI exchange coupling is given by $D_{ij} \cdot (S_i \times S_j)$, which favors spin canting that facilitates the formation of Néel skyrmion.

We have employed the classical atomistic Hamiltonian H to perform the simulation of magnetic textures, as shown below:

$$H = -\frac{1}{2}\sum_{<i,j>} J_{ij} s_i \cdot s_j - \frac{1}{2}\sum_{<i,j>} D_{ij} \cdot (s_i \times s_j) - K_i (s_i \cdot \widehat{K}_i)^2$$
$$-\mu_0 \mu_i H_{ext} \cdot s_i - \mu_0 \mu_i H_{demag} \cdot s_i \quad (1)$$

where $s_i, s_j$ are the normalized spins and $\mu_i, \mu_j$ are the atomic moments at sites $i$, and $j$ respectively. The atomic moment is absorbed into the exchange constant, $J_{ij} = \mu_i \mu_j j_{ij}$, and the DMI interaction $D_{ij} = \mu_i \mu_j d_{ij}$, which is proportional to $r_i \times r_j$, the position vector between the atoms $i$, and $j$ and the interface, and the effective anisotropy $K_i = \mu_i k_i$. $H_{ext}$ and $H_{demag}$ are the external field and demagnetization field respectively. Only the nearest neighbor interactions are considered.

The effective field H$_{eff}$ is calculated using the atomistic Hamiltonian in Eq. (1), and the ground state of the magnetic system is obtained by evolving the spins under the following stochastic Landau-Lifshitz-Gilbert (LLG) equation,

$$\frac{dM}{dt} = -\frac{\gamma}{1+\alpha^2} M \times (H_{eff} + \xi) - \frac{\gamma\alpha}{(1+\alpha^2)M_s} M \times [M \times (H_{eff} + \xi)] \quad (2)$$

where $\gamma$ is the gyromagnetic ratio, $\alpha$ is the Gilbert damping constant, $H_{eff}$ is the effective field, $\xi$ is the Gaussian white noise term for thermal fluctuations and $M_s$ is the saturation magnetization.

**Amorphous Ferrimagnet**

Amorphous rare earth transitional metal (RE-TM) ferrimagnets (FiM) is found to provide a favorable environment to host small skyrmions at room temperature. The magnetic structure of RE-TM FiM consists of two sublattices, one occupied by the RE atoms and the other occupied by the TM atoms. The atoms couple ferromagnetically within each sublattice and antiferromagnetically between the sublattices. The amorphous structure helps to reduce defect pinning, while their intrinsic perpendicular magnetic anisotropy (PMA) allows the formation of Néel skyrmions in thicker films (e.g. up to 10 nm). Furthermore, a distinctive feature of the ferrimagnet is that the magnetization of RE-TM alloys vanishes at the magnetization compensation temperature due to the cancelation of the magnetization of the two sublattices. Figure 3 shows the simulated magnetization of an amorphous Gadolinium-Cobalt ferrimagnet. With near zero magnetization, the skyrmion velocity can reach a high speed near ~1,000 m/s [13]., while near the angular-momentum compensation temperature, the skyrmion Hall effect is vastly reduced [14]. These material advantages make amorphous ferrimagnet an ideal material for spin-based memory and logic devices.

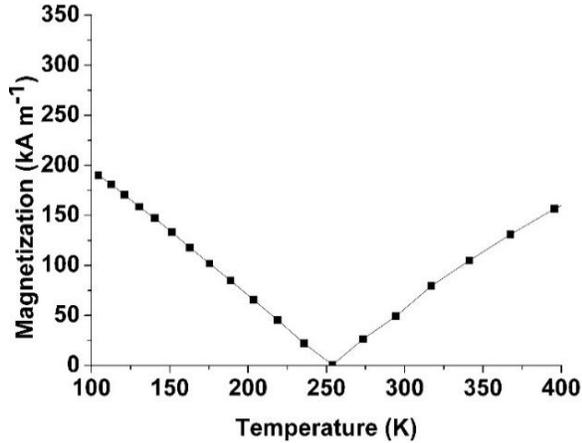

**Fig. 3** Simulated saturation magnetization vs. temperature of amorphous $Gd_{25}Co_{75}$. The compensation temperature is near 250 K, and the magnetization is small at room temperature.

## Results and Discussion

Using atomistic LLG simulations, we will now explore the equilibrium state and size of skyrmion in amorphous $Gd_{25}Co_{75}$ film at room temperature by varying the DMI, magnetic anisotropy, and thickness. To capture the unique short-range order in amorphous materials, an amorphous structure of RE-TM was obtained from *ab initio* molecular dynamics simulation by Professor Howard Sheng using the method described in ref. 15 [15]. The sample size is 50.7 nm x 50.7 nm x 5 nm comprising 768000 atoms. Since the interfacial DMI in these heterostructures originates from the heavy metal interface, we will use an exponential-decay law to describe the DMI inside the magnetic layer. Indeed, such rapid decay of DMI has been found in both calculations [16] and experiments [17].

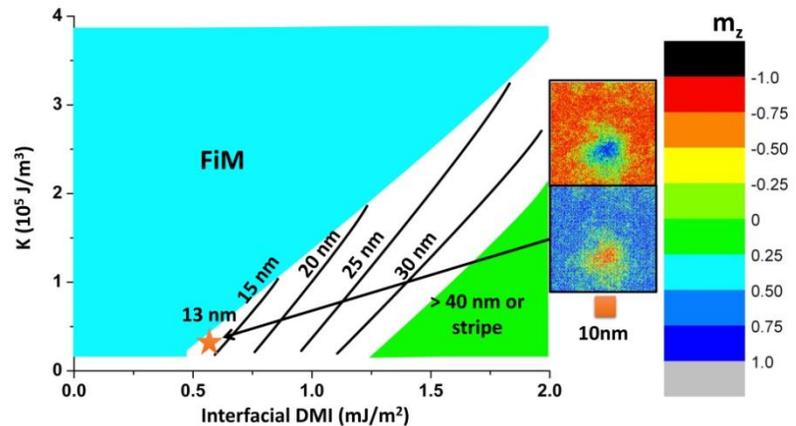

**Fig. 4** Simulated magnetic anisotropy vs. interfacial DMI phase diagram of 5 nm amorphous $Gd_{25}Co_{75}$ at 300 K. Inserted figure corresponds to a 13 nm skyrmions simulated at K = 3 x $10^4$ J/m$^3$. The color maps correspond to Co sublattice (top) and Gd sublattice (bottom).

For a 5-nm thick $Gd_{25}Co_{75}$ layer, interfacial DMI ranges from 0 to 2 mJ/m$^2$ and anisotropy ranges from 0.05 x $10^4$ J/m$^3$ to 4 x $10^5$ J/m$^3$ are investigated. From experiments, the anisotropy of GdCo was found to be ~3 x $10^4$ J/$m^3$ [18] Figure 4 shows the simulated magnetic anisotropy versus interfacial DMI (K-DMI) phase diagram for the 5-nm thick amorphous $Gd_{25}Co_{75}$ at 300 K. For a given anisotropy, as DMI increases from 0 to 2 mJ/$m^2$, the transition from FiM phase to skyrmions, followed by stripe phase occurs. For a given DMI, as the magnetic anisotropy increases, the size of skyrmions decreases, and the skyrmions finally collapse into the FiM state with large enough anisotropy. Using experimental value of K~3 x $10^4$ J/$m^3$, we found skyrmions as small as 13 nm. Such small skyrmion is stabilized with interfacial DMI of ~0.6 mJ/$m^2$. The 2D color map of the out-of-plane reduced magnetization ($m_z$) for the 13 nm skyrmion is inserted in Figure 4. In an experiment, Caretta et al.[13] found skyrmion size in the range of 10 nm to 30 nm in Pt/GdCo/TaO$_x$ with an average DMI of 0.12 mJ/$m^2$, which corresponds to an interfacial DMI of about 0.9 mJ/$m^2$. With such interfacial DMI, our simulation shows a skyrmion size of ~20 nm at room temperature, which is in good agreement with experiment.

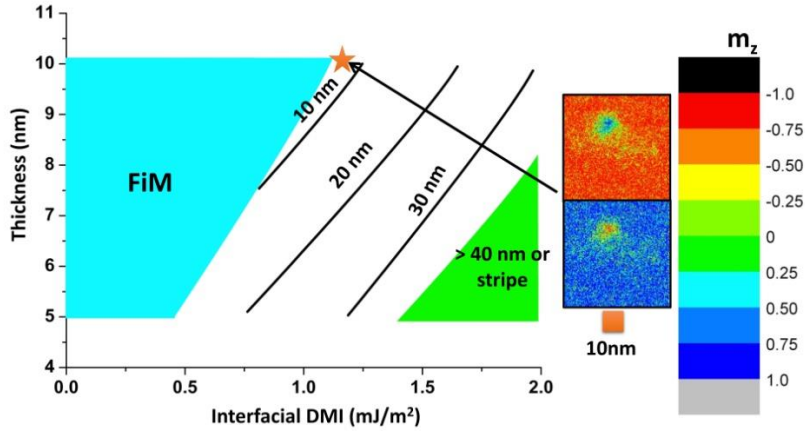

**Fig. 5** Simulated DMI-thickness phase diagram of amorphous $Gd_{25}Co_{75}$ at 300 K with K = 0.3 x $10^5$ J/$m^3$. The inserted figure corresponds to a sub 10 nm skyrmion revealed in 10 nm $Gd_{25}Co_{75}$. The color maps correspond to Co sublattice (top) and Gd sublattice (bottom).

We further investigate the stability of skyrmions at room temperature by increasing the thickness of the GdCo layer. Figure 5 shows the thickness-DMI phase diagram of $Gd_{25}Co_{75}$ at 300 K. For all thicknesses, increase in DMI results in an increase in skyrmion size. The latter can be understood in terms of the effectiveness of DMI in spin canting. Due to the exponential decay of DMI, as thickness increases, the strength of interfacial DMI required to stabilize skyrmions also increases. Even though the interfacial DMI is less effective in the thicker films, smaller skyrmions, as small as 8 nm, are found. In the 10-nm thick GdCo layer, such sub 10 nm skyrmions are stabilized in the DMI range of 1.0 to 1.2 mJ/$m^2$, which is in the range of measured interfacial DMI in Co/Pt films [17]. A color map of reduced magnetization of a sub 10 nm skyrmion is shown in Figure 5. Such skyrmion appears to be robust and contains a well-defined core at the center.

To further demonstrate the robustness of sub 10 nm skyrmions in GdCo, a numerical tomography plot is utilized to reveal the spin texture of the skyrmion in three dimensions. The result is shown in Figure 6. This ultra-small skyrmion is found in the 10-nm GdCo film with an interfacial DMI of 1.1mJ/$m^2$, The color map of $m_Z$ is deliberately set to be brighter in order to more clearly show the skyrmions structure. For the Co sublattice (left of Figure 6), the majority of the spins are pointing down. Near the center, the stripe of green and blue corresponds to the center of the skyrmion. One can conclude from the columnar distribution of green and blue color that this skyrmion is distributed uniformly from top to bottom. Such columnar distribution of skyrmion is also found in the Gd sublattice (right of Figure 6). Such columnar growth of skyrmion is favorable for designing skyrmion-based devices. Spin-based logic devices using these columnar skyrmions will be robust and reliable.

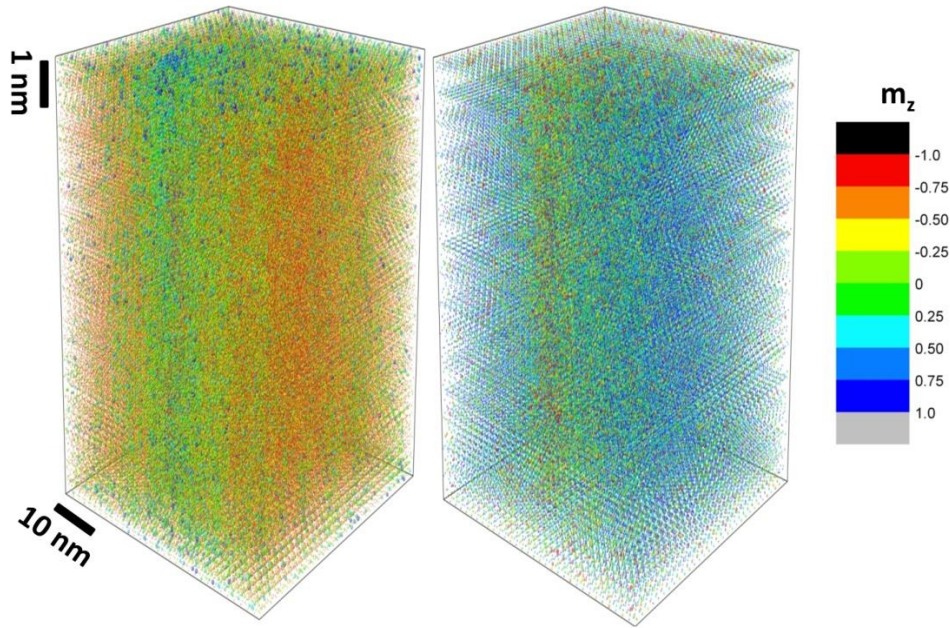

**Fig. 6** Tomograph of a sub 10 nm skyrmion in 10 nm amorphous $Gd_{25}Co_{75}$ at 300 K. Co sublattice is on the left, and Gd sublattice on the right. A robust, columnar distribution of a sub 10 nm skyrmion is revealed in both sublattices.

## Conclusions

Using atomistic simulations, we have explored the phase diagram of Néel skyrmions at room temperature in amorphous ferrimagnetic $Gd_{25}Co_{75}$ films with interfacial Dzyaloshinskii Moriya interaction (DMI). Sub-10 nm skyrmions are found to form in thick (10 nm) films in the range of DMI values similar to that obtained in experiment. Furthermore, despite the exponential decay of DMI away from the interface, 3D spin texture exhibits a uniform columnar distribution across the film thickness. The present study has revealed the robustness of skyrmions, thus adding to the promise of these topological magnetic entities in spin-based nanoelectronics. .

## Acknowledgement:

This work was supported by the DARPA Topological Excitations in Electronics (TEE) program (grant D18AP00009). The content of the information does not necessarily reflect the position or the policy of the Government, and no official endorsement should be inferred. Approved for public release; distribution is unlimited.

# References

1. Geballe, T.H., Matthias, B.T., Clogston, A.M., Williams, H.J., Sherwood, R.C., Maita, J.P.: Localized moments (?). J. Appl. Phys. **37**, 1181–1186 (1966)
2. Wolf, S. A., Chtchelkanova, A. Y., Trege, D. M.: Spintronics—a retrospective and perspective. IBM J. Res. Dev. **50** (1), 101-110 (2006)
3. Parkin, S., Yang, S.-H.: Memory on the racetrack. Nat. Nanotechnol. **10**, 195-198 (2015)
4. Kang, W., Huang, Y., Zhang, X., Zhou, Y., Zhao, W.: Skyrmion-electronics: an overview and outlook. Proc. IEEE **104** (10), 2040-2061 (2016)
5. Sampaio, J., Cros, V., Rohart, S., Thiaville, A., Fert, A.: Nucleation, stability and current-induced motion of isolated magnetic skyrmions in nanostructures. Nat. Nanotechnol. **8** 839-844 (2013)
6. Dzyaloshinsky, I.: A thermodynamic theory of weak ferromagnetism of antiferromagnetics. J. Phys. Chem. Solids **4**, 241–255 (1958)
7. Moriya, T.: Anisotropic superexchange interaction and weak ferromagnetism. Phys. Rev. **120**, 91–98 (1960)
8. Mühlbauer, S., Binz, B., Jonietz, F., Pfleiderer, C., Rosch, A., Neubauer, A., Georgii, R. & Böni, P.: Skyrmion lattice in a chiral magnet. Science **323**, 915-919 (2009).
9. Yu, X.Z., Kanazawa, N., Onose, Y., Kimoto, K., Zhang, W. Z., Ishiwata, S., Matsui, Y. Tokura, Y.: Near room-temperature formation of a skyrmion crystal in thin-films of the helimagnet FeGe. Nat. Mater. **10**, 106–109 (2011)
10. Tolley, R., Montoya, S.A., Fullerton, E. E.: Room-temperature observation and current control of skyrmions in Pt/Co/Os/Pt thin films. Phys. Rev. Mater. **2**, 044404 (2018)
11. Woo, S., Litzius, K., Krüger, B., Im, M.-Y., Caretta, L., Richter, K., Mann, M., Krone, A., Reeve, R. M., Weigand, M., Agrawal, P., Lemesh, I., Mawass, M.-A., Fischer, P., Kläui, M., Beach, G.S.D.: Observation of room-temperature magnetic skyrmions and their current-driven dynamics in ultrathin metallic ferromagnets. Nat. Mater. **15**, 501–506 (2016)
12. Soumyanarayanan, Raju, A. M., Gonzalez Oyarce, A. L., Tan, A. K. C., Im, M.-Y., Petrović, A. P., Ho, P., Khoo, K. H., Tran, M., Gan, C. K., Ernult, F., Panagopoulos, C.: Tunable room-temperature magnetic skyrmions in Ir/Fe/Co/Pt multilayers. Nat. Mater. **16**, 898–904 (2017)
13. Caretta, L., Mann, M., Büttner, F., Ueda, K., Pfau, B., Günther, C. M., Hessing, P., Churikova, A., Klose, C., Schneider, M., Engel, D., Marcus, C., Bono, D., Bagschik, K., Eisebitt, S., Beach, G. S. D.: Fast current-driven domain walls and small skyrmions in a compensated ferrimagnet. Nat. Nanotechnol. **13**, 1154-1160 (2018)
14. Woo, S., Song, K. M., Zhang, X., Zhou, Y., Ezawa, M., Liu, X., Finizio, S., Raabe, J., Lee, N. J., Kim, S.-I., Park, S.-Y., Kim, Y., Kim, J.-Y., Lee, D., Lee, O., Choi, J. W., Min, B.-C., Koo, H. C., Chang, J.: Current-driven dynamics and inhibition of the skyrmion Hall effect of ferrimagnetic skyrmions in GdFeCo films. Nat. Commun. **9**, 959 (2018)
15. Sheng, H.W., Luo, W.K., Alamgir, F.M., Bai, J.M., Ma, E.: Atomic packing and short-to-medium-range order in metallic glasses. Nature **439**, 419-425 (2006)
16. Yang, H., Thiaville, A., Rohart, S., Fert, A., Chshiev, M.: Anatomy of Dzyaloshinskii-Moriya interaction at Co/Pt interfaces. Phys. Rev. Lett. **118**, 219901 (2017)
17. Stashkevich, A. A, Belmeguenai, M., Roussigné, Y., Cherif, S. M., Kostylev, M., Gabor, M., Lacour, D., Tiusan, C., Hehn, M.: Experimental study of spin-wave dispersion in Py/Pt film structures in the presence of an interface Dzyaloshinskii-Moriya interaction. Phys. Rev. B **91**, 214409 (2015)
18. Deng, M., Poon, S. J.: Tunable perpendicular magnetic anisotropy in GdFeCo amorphous films. J. Magn. Magn. Mater. **339**, 51-55 (2013)